\documentclass[aps,prl,twocolumn,longbibliography,superscriptaddress]{revtex4-1}

\usepackage{amsmath,amssymb,amsthm,amsfonts,mathrsfs,dsfont,braket,array,float}
\usepackage{color,xcolor,graphicx}
\usepackage{slashed}
\usepackage{enumitem}
\usepackage{stmaryrd} %extra symbols
\usepackage{esint}
\usepackage{physics}
\usepackage{braket}
\usepackage{tensor}
\usepackage{float}
\usepackage{mathtools}
\usepackage{bbold}
\usepackage{shortcuts}
\usepackage{microtype}   % improves spacing

\definecolor{rossocorsa}{rgb}{0.83, 0.0, 0.0}
\definecolor{navyblue}{rgb}{0.0, 0.0, 0.5}
\definecolor{custom}{rgb}{0.05,0.31,0.55}

\usepackage[
  bookmarks=true,
  bookmarksnumbered=false,
  hyperindex=true,
  bookmarksopen=true,
  hyperfigures=true,
  colorlinks=true,
  linkcolor=navyblue,
  citecolor=rossocorsa,
  urlcolor=custom,
  breaklinks]{hyperref}

\begin{document}

\title{ER for typical EPR}

\author{Javier M.~Mag\'{a}n}
\affiliation{Instituto Balseiro, Centro At\'omico Bariloche, 8400-S.C. de Bariloche, R\'io Negro, Argentina}

\author{Martin~Sasieta}
\affiliation{Martin Fisher School of Physics, Brandeis University, Waltham, Massachusetts 02453, USA}
\author{Brian~Swingle}
\affiliation{Martin Fisher School of Physics, Brandeis University, Waltham, Massachusetts 02453, USA}

\begin{abstract}
What do the typical entangled states of two black holes look like? Do they contain semiclassical interiors? We approach these questions constructively, providing ensembles of states which densely explore the black hole Hilbert space. The states contain very long {\it Einstein-Rosen (ER) caterpillars}: semiclassical wormholes with large numbers of matter inhomogeneities.  Distinguishing these ensembles from the typical entangled states of the black holes is hard. We quantify this by deriving the correspondence between a microscopic notion of quantum randomness and the geometric length of the wormhole. This formalizes a ``complexity = geometry'' relation.  
\end{abstract}

\maketitle

\textbf{Introduction} --- ER=EPR postulates the remarkable equivalence between spatial wormholes in general relativity (Einstein-Rosen bridges, ER) and quantum entanglement (Einstein-Podolsky-Rosen, EPR) \cite{Maldacena:2013xja}. In essence, when two systems are quantum-mechanically entangled, a “wormhole” is conjectured to be connecting them.

The proposal is fundamentally rooted in holographic entanglement in the Anti-de Sitter space / conformal field theory correspondence (AdS/CFT) \cite{Ryu:2006bv}. There, the observation is that an honest geometric wormhole can emerge between  two black holes when the microscopic quantum correlations are large enough \cite{VanRaamsdonk:2010pw}. The classic example is the thermofield-double state (TFD) of two copies of the holographic CFT on $\mathbf{R}\times \mathbf{S}^{d-1}$,
\be\label{eq:TFD}
\ket{\text{TFD}} = \frac{1}{\sqrt{Z(\beta)}} \sum_{i} e^{-\frac{\beta}{2} E_i} \ket{E_i^*}_L\ket{E_i}_R\,,
\ee
where $|E_i\rangle_{L,R}$ are energy eigenstates of the left ($L$) and right ($R$) CFT Hamiltonians $H_{L,R}$. The star in $\ket{E_i^*}$ represents the action under $\mathsf{CRT}$. On one hand, this is an entangled state between the CFTs which canonically purifies the thermal state of a single CFT. On the other hand, the TFD is dual to a connected spatial wormhole, corresponding to the Hartle–Hawking state of an eternal black hole in AdS space \cite{Maldacena:2001kr}.

The TFD example is very illuminating but, as often emphasized \cite{Maldacena:2013xja,Maldacena:2001kr,Shenker:2013pqa,Shenker:2013yza,Marolf:2013dba}, it is rather atypical, given that the TFD is a very special entangled state. For one thing, it lies in the diagonal subspace, annihilated by $H_L - H_R$. Additionally, its wavefunction has large correlations between the two systems, captured geometrically by a short wormhole with a bifurcate Killing horizon. Such fine-tuned correlations are dynamically destroyed by small perturbations injected a scrambling time in the past \cite{Shenker:2013pqa}.

If ER = EPR is to represent a general principle, one might ask: is there a wormhole for a typical entangled state of two black holes?

\textbf{Typical EPR} --- To motivate a definition of ``typical EPR'' that makes sense for CFTs in thermal equilibrium, we first consider a toy example consisting of two collections of $N$ qubits, prepared in the maximally entangled state
\be\label{eq:inftempTFD}
\ket{I} = \dfrac{1}{\sqrt{D}}\sum_{i=1}^D \ket{i^*}_L\ket{i}_R\,,
\ee 
where the $\ket{i}\equiv \ket{i_1}\otimes ...\otimes \ket{i_N}$ form an orthonormal basis and the dimension of each Hilbert space is $D = 2^N$. The state \eqref{eq:inftempTFD} is the infinite-temperature TFD, in which the systems share $N$ Bell pairs.

We will define the ensemble of {\it typical entangled states} $\ens_{\text{EPR}}$ between these systems, as the collection of states obtained by applying a random single-sided unitary to the TFD, $\ket{\Psi_{\text{EPR}}} =   U_{R}\ket{I} $, or more explicitly,
\be\label{eq:randqubit} 
\ket{\Psi_{\text{EPR}}} =    \dfrac{1}{\sqrt{D}}\sum_{i,j=1}^D U_{ij} \ket{i^*}_L\ket{j}_R\,,
\ee 
where $U$ is a Haar random unitary in $\U(D)$. The state \eqref{eq:randqubit} is a random purification of the maximally mixed state $\rho_L = \rho_R = \frac{1}{D}\mathbf{1}$.

Now, to define the typical entangled states of two black holes, we will proceed by analogy and generalize the notion above to finite temperature. In AdS/CFT, an equilibrium black hole has an associated equilibrium density matrix $\rho$ of the CFT. For a neutral black hole, the state $\rho$ is ensemble-equivalent to the canonical Gibbs state $\rho_\beta = e^{-\beta H}/Z(\beta)$ in the large-$N$ limit of the CFT. 

We define the ensemble of {\it typical equilibrium entangled states} $\ens^\rho_{\text{EPR}}$ as the collection of states of the form $(\sqrt{\rho}\otimes \sqrt{\rho})\ket{\Psi_{\text{EPR}}}$, or explicitly
\be\label{eq:randBH}
\ket{\Psi^\rho_{\text{EPR}}} \,\propto\, \sum_{i,j} \sqrt{\rho_i \rho_j}\, X_{ij}\ket{E_i^*}_L\ket{E_j}_R\,.
\ee
Here $\rho_i$ is the eigenvalue of $\rho$ in the state $\ket{E_i}$ and we have introduced $X = \sqrt{D}\,U$ for Haar random $U \in \U(D)$. With this normalization, we can safely take the infinite Hilbert space dimension limit $D\rightarrow \infty$. In this limit the coefficients $X_{ij}$ converge to i.i.d. Gaussian random variables with zero mean and unit variance. In appendix A, we outline some of the properties of the ensemble $\ens^\rho_{\text{EPR}}$. For example, on average, the reduced state is the equilibrium density matrix, $\mathbb{E}\left[\rho_{L,R}\right] = \rho$.

Our question becomes: do such typical equilibrium entangled states have wormholes associated with them? Our approach in this letter is to explicitly construct ensembles of states with wormholes that better and better approximate such typical EPRs. 

\textbf{A dense exploration of Hilbert space} --- Random states and unitaries are ubiquitous in quantum physics, but implementing them in practice is computationally hard. In the present context, the objective is to, even in principle, construct a typical-looking entangled state of two black holes starting from the TFD.

A natural possibility is to consider the time evolution of the TFD state $\ket{\text{TFD}_t} = e^{-\iw t H_R}\ket{\text{TFD}}$, or explicitly,
\be\label{eq:TDFt}
\ket{\text{TFD}_t} = \frac{1}{\sqrt{Z(\beta)}} \sum_{i} e^{-\left(\frac{\beta}{2} + \iw t\right) E_i} \ket{E_i^*}_L\ket{E_i}_R\,.
\ee
As we elaborate in appendix B, $\ket{\text{TFD}_t}$ will never become a random EPR, even if the Hamiltonian is completely structureless. The main reason is the conservation of energy along the trajectory, which prevents the state vector from exploring most of Hilbert space.

This example nevertheless provides a more general instance of ER = EPR. From the spacetime perspective, the wormhole grows in time \cite{Maldacena:2013xja,Hartman:2013qma}. The geometric volume of the wormhole is conjectured to encode a suitable notion of ``quantum complexity'' of the time-evolution \cite{Susskind:2014rva}.

The way to continuously reach random-looking entangled states is to instead consider time-dependent Hamiltonians driving the evolution. For the sake of clarity, we shall first describe this in the qubit system. We will consider a general time-dependent Hamiltonian of the form
\be\label{eq:timedepH} 
H(t) =  \sum_{\alpha =1}^K g_\alpha(t) \mathcal{O}_\alpha\,,
\ee
where the $ 
\mathcal{O}_\alpha$ form a collection of $K$ Hermitian operators, normalized so that $\text{Tr}(\Op_\alpha^2)=D$, with associated time-dependent couplings $ 
g_\alpha(t)$. The time-evolution operator now becomes
\be 
U(t) = \mathbf{T}\left\lbrace \exp\left(-\iw \int_0^t \text{d}s \,H(s)\right)\right\rbrace \,.
\ee 
Applying it to the TFD generates states $\ket{\Psi_t}  = U(t)_R\ket{I}$, or explicitly,
\be\label{eq:inftempRC}
\ket{\Psi_t}  = \dfrac{1}{\sqrt{D}}\sum_{i,j=1}^D U(t)_{ij}\ket{i^*}_L\ket{j}_R\,.
\ee 

In this letter, we will select the couplings from a collection of white-noise correlated Gaussian random couplings
\be\label{eq:statisticscouplings} 
\mathbb{E}\left[g_\alpha(t)\right] = 0\,,\quad \mathbb{E}\left[g_\alpha(t)g_{\alpha'}(t')\right] = J\delta_{\alpha\alpha'} \delta(t-t')\,.
\ee
Such couplings define an independent random Hamiltonian at each time step. Upon exponentiation, $U(t)$ follows a multiplicative Brownian motion in $\U(D)$, which corresponds to the continuous-time analog of a random quantum circuit. In particular,  at each time-step, the time-evolution implements an infinitesimal ``random gate'' $\exp(-\iw\delta t\,  H(t))$. At the level of the states \eqref{eq:inftempRC}, this Markov process defines an ensemble of states $\ens_t$. 

Under general assumptions for the drive operators, and in the absence of symmetries, it is possible to show that the measure of the ensemble of states $\ens_t$ converges weakly to the uniform measure at infinite times \cite{Jian:2022pvj,Guo:2024zmr}
\be
\lim\limits_{t\rightarrow \infty} \ens_t = \ens_{\text{EPR}}\,,
\ee 
or, equivalently, that $U(\infty)$ is Haar random.

\begin{figure}[h]
    \centering
    \includegraphics[width=0.65\linewidth]{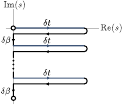}
    \caption{Complex time contour $\Gamma_t$ defining the operator $U(\Gamma_t)$. The Hamiltonian $H_c(s)$ is taken to be $H_0$ on the black parts of the contour and $H_0 +  H(s)$ on the blue parts. The timefolds effectively implement the interaction Hamiltonian $H(s)$.}
    \label{fig:contour}
\end{figure}

\textbf{Gradual cooling} --- For black holes, we will consider a holographic random quantum circuit selecting a collection of perturbations $\Op_\alpha$ of the CFT associated to low-energy matter fields in the bulk. An obvious problem which arises here is that applying $U(t)$ to a finite-temperature TFD will eventually heat up the corresponding black hole, and this process will not generate a typical equilibrium EPR of the form \eqref{eq:TDFt}. To stabilize the energy of the black hole, we will instead gradually cool down the random circuit applying the operator \cite{Magan:2024aet}
\be 
U(\Gamma_t) = e^{-\delta \beta H_0} U(n,\delta t) e^{-\delta \beta H_0} \cdots e^{-\delta \beta H_0} U(1,\delta t) \,.
\ee
where $n = t/\delta t$. Formally, this defines an operator
\be 
U(\Gamma_t) = \mathbf{P}\left\lbrace \exp\left(-\iw \int_{\Gamma_t} \text{d}s \,H_c(s)\right)\right\rbrace
\ee 
by path-ordering on a contour $\Gamma_t$ shown in Fig. \ref{fig:contour}. We follow an infinitesimal cooling-and-evolving process, with $\delta t= \delta \beta \rightarrow 0$, where the relative magnitude between $\delta t$ and $\delta \beta$ has been absorbed into the definition of $J$ in \eqref{eq:statisticscouplings}. 

We will utilize the gradually cooled random circuit to define the ensemble of states $\ens_{\Gamma_t}$ at fixed time, 
\be\label{eq:fintempRC}
\ket{\Psi_{\Gamma_t}} \, \propto\, \sum_{i,j} e^{-\frac{\beta}{4} (E_i  +E_j)}U(\Gamma_t)_{ij}\ket{E_i^*}_L\ket{E_j}_R\,.
\ee 
At $t=0$ the state \eqref{eq:fintempRC} corresponds to the TFD \eqref{eq:TFD}. As we will later explain, when $t\rightarrow \infty $, $U(\Gamma_\infty )$ converges weakly to a random matrix $\sqrt{\rho_0} \,X \sqrt{\rho_0}$ for Gaussian random $X$ and a density matrix $\rho_0$. The state $\rho_0$ is determined by the steady state of the random quantum circuit \cite{Magan:2024aet}. Accordingly, the ensemble $\ens_{\Gamma_t}$ converges to the ensemble of typical entangled states of the two black holes \be\label{eq:convergencefinite}
\lim\limits_{t\rightarrow \infty} \ens_{\Gamma_t} = \ens^\rho_{\text{EPR}}\,,
\ee 
for the equilibrium state 
\be\label{eq:equilibriumstate} 
\rho = e^{-\frac{\beta}{4}H_0} \rho_0e^{-\frac{\beta}{4}H_0}\,.
\ee

Thus, the random circuit allows us to continuously connect the TFD and a typical EPR of two black holes.

\begin{figure}[H]
    \centering
    \includegraphics[width=0.8\linewidth]{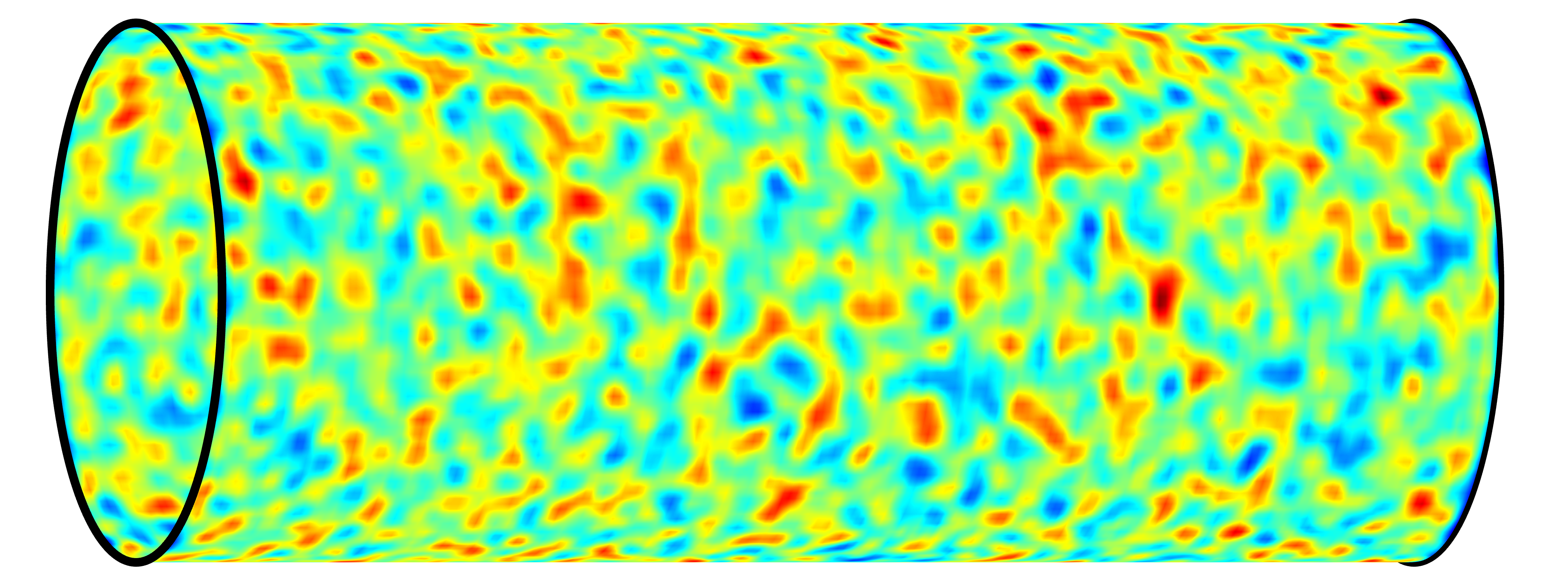}
    \caption{The ER caterpillar is a long bumpy wormhole supported by an inhomogeneous matter distribution, with correlation scale set by $\ell_{\Delta}$ and average length set by $\ell(t)$.}
    \label{fig:caterpillarinstance}
\end{figure}

\textbf{ER caterpillars} --- We will now elucidate geometric properties of the semiclassical duals to $\ket{\Psi_{\Gamma_t}}$, using the gravitational path integral to prepare them. While the precise geometry of individual states will be difficult to characterize, we will argue that they include two black holes connected by an Einstein-Rosen (ER) caterpillar~\footnote{To our knowledge, related sorts of caterpillars were first discussed by H. Lin and L. Susskind in the different context of exponentially long wormholes arising from a fixed Hamiltonian evolution \cite{LennyTalk,Susskind:2020wwe}}: a spatial wormhole with a large number of matter inhomogeneities and geometric features qualitatively similar to those illustrated in Fig.~\ref{fig:caterpillarinstance}.

\vspace{-.5cm}

Unlike the multiple-shock spacetimes of \cite{Shenker:2013yza}, the caterpillars are not spherically symmetric and are constructed by introducing matter from the Euclidean section directly into the black hole interior, thereby preserving equilibrium and allowing the wormhole to become much longer.

\begin{figure}[h]
    \centering
    \includegraphics[width=0.87\linewidth]{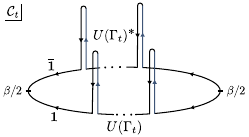}
    \caption{Complex time contour $\mathcal{C}_t$ composed of forward ($1$) and backward ($\bar{1}$) contours.}
    \label{fig:contour2}
\end{figure}

In order to construct them, we start from the norm of the right-hand side of \eqref{eq:fintempRC}, which is proportional to the correlation function $Z_{\Psi}(t) =  Z(\beta)^{-1}\text{Tr}(e^{-\frac{\beta}{2}H_0}U(\Gamma_t) e^{-\frac{\beta}{2}H_0}U({\Gamma}_t)^\dagger)$. This quantity can be alternatively expressed as the survival amplitude \cite{Magan:2024aet}
\be\label{eq:normoverlap} 
Z_{\Psi}(t) =  \bra{\TFD} U(\Gamma_t)_1 \otimes U(\Gamma_t)_{\bar{1}}^*\ket{\TFD}\,.
\ee 
Here we have introduced the TFD state between forward $(1)$ and backward $(\bar{1})$ contours, at inverse temperature $\beta$. In this form, $Z_\Psi(t)$ is computed by the CFT path integral along a complex time-contour $\mathcal{C}_t$ shown in Fig. \ref{fig:contour2}.  The time-orientation of the backward contour is reversed, as $U(\Gamma_t)_{\bar{1}}^*$ is applied on this contour. The unnormalized state \eqref{eq:fintempRC} is prepared by the lower half of this contour.

Using the AdS/CFT dictionary, $Z_\Psi(t)$ can be evaluated semiclassically by the gravitational path integral with sources for the bulk fields $g_\alpha(t)$ given by the time-dependent couplings of $U(\Gamma_t)$. For single instances of the circuit, evaluating the gravitational path integral in a saddle point approximation is highly non-trivial. A trick that we will use is to average over the time-dependent couplings and evaluate the average norm
\be\label{eq:avnorm}
Z(t) \equiv \expectation\left[Z_{\Psi}(t)\right]\,.
\ee 
Given that the perturbations that we have chosen are semiclassical for single realizations of the couplings, $Z(t)$ will give us information about the average geometry over the ensemble of states. This is justified because the norm of the states --and thus the semiclassical geometry that computes it-- is self-averaging over $\ens_{\Gamma_t}$ \cite{Magan:2024aet}. At late times, $\text{Std}\left[Z_{\Psi}(t)\right]/Z(t) \sim e^{-S_2(\rho)}$ where $S_2(\rho) = -\log \text{Tr}(\rho^2)$ is the second R\'{e}nyi entropy of the equilibrium state \eqref{eq:equilibriumstate}.

The average over the Brownian couplings can be performed exactly, using the identity 
\be\label{eq:identitybrownian} 
\expectation\left[U(\Gamma_t)_1 \otimes U(\Gamma_t)^*_{\bar{1}}\right] = \exp(- t H_{\eff,1})\,,
\ee 
for the effective time-independent Hamiltonian
\be\label{eq:effHgen} 
H_{\eff,1} = H_0^1 + H_0^{\bar{1}}   + \dfrac{J}{2 }  \sum_{\alpha=1}^K \left(\Op_\alpha^{1}-\Op_\alpha^{\bar{1}}\,^*\right)^2\,.
\ee 
That is, the disorder average over the Brownian couplings effectively produces local-in-time interactions between both contours in the form of a time-independent Hamiltonian $H_{\eff,1}$. The quadratic interaction in \eqref{eq:effHgen} arises from the Gaussianity of the random couplings.

Inserting the identity \eqref{eq:identitybrownian} into \eqref{eq:avnorm} we find that the average norm corresponds to the matrix element
\be\label{eq:averagenormmatrixelement}
Z(t) = \bra{\TFD}\exp(-tH_{\eff,1})\ket{\TFD}\,.
\ee 

This is considerably simpler because \eqref{eq:averagenormmatrixelement} is a purely Euclidean object, and so is the gravitational saddle point manifold $M_t$ that evaluates it semiclassically,
\be\label{eq:normcaterbulkav} 
Z(t) \sim  Z_{\text{grav}}[M_t] \quad\quad \text{as } N\rightarrow \infty\,,
\ee
where $Z_{\text{grav}}[M_t] = e^{-I_{\text{grav}}[M_t]}Z_{1\text{-loop}}$, with $I_{\text{grav}}[M_t]$ the Euclidean gravitational action and $Z_{1\text{-loop}}$ the one-loop partition function for the bulk fields on $M_t$. 

The relevant properties of $M_t$ follow from an assumption for the effective Hamiltonian of the random circuit:

{\it At large $N$, the effective Hamiltonian}  $H_{\eff,1}$ {\it \eqref{eq:effHgen} is gapped and it contains a unique ground state} $\ket{\text{GS}}$ {\it with a semiclassical description as a connected spatial wormhole.} 

\begin{figure}[H]
    \centering
    \includegraphics[width=\linewidth]{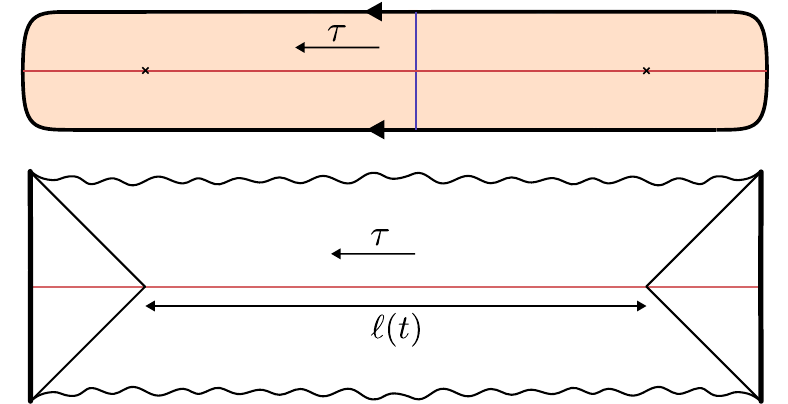}
    \caption{On top, the Euclidean saddle point $M_t$ of topology $D \times \mathbf{S}^{d-1}$, where $D$ is a disk, has an approximate translation isometry in the $\tau$ direction. At each constant-$\tau$ (blue) slice the path integral prepares the semiclassical dual to $\ket{\text{GS}}$. On the bottom, the Lorentzian continuation across the red slice preparing a two-sided black hole with an approximate translation symmetry in the interior. The length of the wormhole $\ell(t)$ is proportional to the circuit time $t-t_\star$.}
    \label{fig:caterpi}
\end{figure}

We will later motivate specific choices of the drive operators $\Op_\alpha$ which lead to an effective Hamiltonian satisfying this assumption, but keep the argument general here. Under the assumption above, the operator $\exp(-sH_{\eff,1})$ is approximately proportional to the projector $\ket{\text{GS}}\bra{\text{GS}}$ for $s\geq t_\star$. What this means for $Z(t)$ is that there is a Euclidean time-translation symmetry of the CFT path integral $Z(t) \approx \bra{\TFD}\exp(-s H_{\eff,1})\ket{\text{GS}}\bra{\text{GS}}\exp(-(t-s) H_{\eff,1})\ket{\TFD}$ and the bulk saddle point $M_t$ must inherit this isometry. The onset time of this symmetry is $t_\star \sim \Egap^{-1} \log N_*$, where $\Egap$ is the gap and $N_*$ is the number of first excited states of $H_{\eff,1}$. As illustrated in Fig. \ref{fig:caterpi}, this symmetry deforms the geometry of the disk and makes it very long. Cutting the path integral open at a constant Euclidean time prepares the unnormalized ground state, represented in the bulk as a connected spatial wormhole (blue slice). For this reason, the metric on $M_t$ is that of a Euclidean eternal traversable wormhole \cite{Magan:2024aet}. The relevant slice that determines the average geometry of the ensemble of ER caterpillars is the red reflection-symmetric slice on $M_t$. Because of the isometry, this slice contains an approximately cylindrical spatial wormhole of length scaling linearly with $t-t_\star$.

Since the effective Hamiltonian is gapped, the lightest particle excitations on this wormhole will correspond to the first excited states, with energy $\Egap$. These excitations serve as a bulk ``clock'' to measure the wormhole length $\ell(t)$ and relate it to the asymptotic circuit time. The precise relation is \cite{Magan:2024aet}
\be\label{eq:length} 
\dfrac{\ell(t)}{\ell_{\Delta}} = {\Egap}  \left(t-t_\star\right)\,,
\ee 
where $\ell_{\Delta}$ is the correlation scale of the lightest excitations on the wormhole. The quantities $\Egap$ and $\ell_\Delta$  are determined by quantizing the bulk fields on $M_t$, with suitable boundary conditions, and reading their energy and decay of the bulk two-point function.

For explicit constructions of the ER caterpillars, we could specify the drive operators $\Op_\alpha$ to be local single-trace relevant conformal primaries, in which case $\alpha$ labels spatial points. In this case, the interaction \eqref{eq:effHgen} corresponds to an eternal version of the Gao-Jafferis-Wall double-trace perturbation \cite{Gao:2016bin}. Coincidentally, the effective Hamiltonian \eqref{eq:effHgen} was analyzed in detail in the SYK model by Maldacena and Qi, with an interaction term formed by a large number $K\sim N$ of relevant Majorana bilinears \cite{Maldacena:2018lmt}. In that case, the lesson is that the low-temperature phase of the coupled systems is gapped, with a unique ground state that corresponds to a connected spatial wormhole, satisfying our assumptions. The ground state is similar to the TFD at some coupling-dependent temperature $\beta(J)$ (in units of the SYK couplings). Moreover, it has $O(1)$ overlap with it for large-$q$ SYK and for $1/N \ll J \ll 1$ in the regular $q=4$ SYK (in units where the SYK couplings $\mathcal{J}=1$). In \cite{Magan:2024aet} we provided a totally explicit construction of the ER caterpillars of the SYK model based on this construction. A similar picture is expected to hold in higher dimensions, provided that the Hamiltonian $H_0$ is sufficiently chaotic \cite{Cottrell:2018ash}. 

\textbf{Wormhole length = Randomness} --- We will now derive a quantitative relation between the average length of the wormhole $\ell(t)$ and a microscopic feature of the ensemble of states $\ens_{\Gamma_t}$. Let us explain this notion in the qubit system first. The idea is that, at finite but large circuit time $t$, the states \eqref{eq:inftempRC} are not fully typical EPRs because $\ens_t$ and $\ens_{\text{EPR}}$ are statistically distinguishable. The approach to typical is quantified by the notion of a {\it quantum state $k$-design}: an ensemble of states which reproduces the first $k$ moments of $\ens_{\text{EPR}}$. More precisely, the  $k$-th moment superstate of a general ensemble of states $\ens_{\Psi}$ is defined as
\be\label{eq:momentsgen}
\rho_k(\ens_{\Psi}) = \expectation_{\ens_\Psi}[ \left(\ket{\Psi} \bra{\Psi}\right)^{\otimes k}]\,.
\ee 
The state $\rho_k(\ens_{\Psi})$ is defined on $2k$ replicas of the original Hilbert space. It is generally a mixed state, given the correlations between replicas that arise from the average over $\ens_{\Psi}$. Given this, a quantum state $k$-design is defined by the condition $\rho_k(\ens_{\Psi}) = \rho_k(\ens_{\text{EPR}})$. For $\ens_{\Psi} = \ens_t$, this is equivalent to the condition that the random circuit $U(t)$ forms a {\it unitary $k$-design}: an ensemble of unitaries which reproduces the first $k$ moments of the Haar ensemble. 

Note that these notions are physically very strong, since a $k$-design is indistinguishable from $\ens_{\text{EPR}}$ by any statistical $k$-copy measure. In practice, only approximate notions of $k$-designs make sense at finite time. For these, one typically bounds some quantum distance between moment superstates. 

We define a ($\rho$ equilibrium) quantum state $k$-design by the condition $\rho_k(\ens_{\Psi}) = \rho_k(\ens^\rho_{\text{EPR}})$. For $\ens_{\Psi} = \ens_{\Gamma_t}$ we will compute the distance to $k$-design
\be\label{eq:deltadef2} 
\Delta^\rho_{k}(t) \equiv \dfrac{\big\|\rho_k(\ens_{\Gamma_t}) - \rho_k(\ens^\rho_{\text{EPR}})\big\|_{2}}{\big\|\rho_1(\ens^\rho_{\text{EPR}})\big\|_{2}^k} \,,
\ee 
and the corresponding time to $k$-design $t_{k} = \min\limits\left\lbrace  t:\Delta^\rho_{k}(t)\leq  \varepsilon\right\rbrace$. We are now ready to state the main result of this letter:\\[.1cm]

{\bf Length-randomness correspondence.} {\it The ensemble of ER caterpillars of average length $\ell$ and matter correlation scale $\ell_{\Delta}$  forms an $\varepsilon$-approximate quantum state $k$-design of the black holes for  
\be\label{eq:length_randomness}
k \sim  \dfrac{\ell - \ell_\varepsilon}{\ell_{\Delta}}\,,
\ee
where $\ell_\varepsilon = \ell_{\Delta} \log \varepsilon^{-1}$ and $k \ll O(e^{S_2(\rho)})$.
}\\

The correspondence follows from evaluating the distance $\Delta^\rho_{k}(t)$ using the gravitational path integral. The calculation boils down to gravitationally computing the {\it $k$-th moment two-point function} of the ensemble
\be\label{eq:FPthermal} 
G_k(t,t') \equiv \text{Tr}\left(\rho_k(\ens_{\Gamma_t})\rho_k(\ens_{\Gamma_{t'}})\right) = \expectation|\bra{\Phi_{\Gamma_{t'}}}\ket{\Psi_{\Gamma_t}} |^{2k}\,.
\ee 
This quantity corresponds to the $2k$-th moment of the overlap between two independent draws of the ensemble of states. In the CFT, it is evaluated as a disordered path integral on $2k$ replicas of a closed time contour, with two independent instances of the random circuit on each contour, generalizing Fig.~\ref{fig:contour2}. The moment two-point function is relevant because it controls the distance to design
\be\label{eq:2norm} 
\Delta^\rho_{k}(t)^2 = \dfrac{F_k(t) - 2G_k(t,\infty) + F_k(\infty)}{F_1(\infty)^{k}}\,,
\ee 
where we have defined the purity of the moment state $F_k(t) = G_k(t,t)$, also dubbed the $k$-th frame potential.

\begin{figure}[H]
    \centering
    \includegraphics[width=\linewidth]{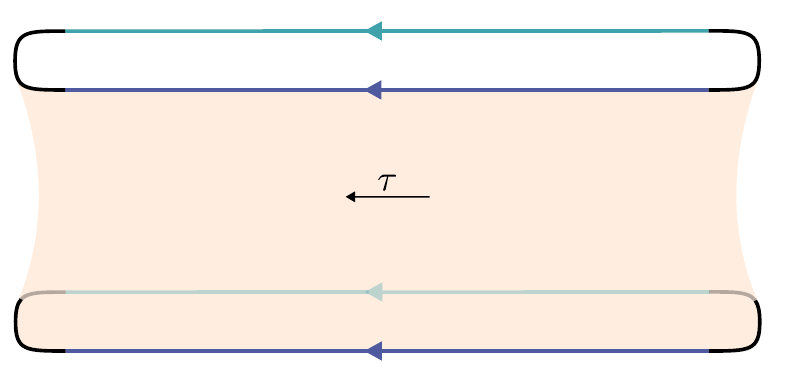}
    \caption{Euclidean two-boundary wormhole providing the late-time ``plateau'' value of $G_1(t,t')$. The saddle-point topology is $\mathbf{S}^1\times \mathbf{R}\times \mathbf{S}^{d-1}$. The wormhole geometry includes two long regions with a Euclidean time $\tau$ translation isometry geometrically equivalent to the one for the disk in Fig. \ref{fig:caterpi}.}
    \label{fig:cosmo}
\end{figure}

Now, the average over Brownian couplings in \eqref{eq:FPthermal} can be performed exactly. This produces time-independent interactions between the $2k$ contours, and the moment two-point function corresponds to a thermal-looking partition function
\be\label{eq:genFP} 
G_k(t,t') \,\propto\, \text{Tr}\left(e^{-\frac{\beta}{2}H_{0,k}}e^{-t H_{\eff,k}} e^{-\frac{\beta}{2}H_{0,k}}e^{-t' H_{\eff,k}}\right) \,.
\ee 
for the time-independent $2k$-replica Hamiltonian
\be\label{eq:effHkfinitetemp} 
H_{\eff,k} =  H_{0,k}  + \dfrac{J}{2} \sum_{\alpha=1}^K \left(\sum_{r=1}^k \Op_\alpha^{r} - \Op_\alpha^{\bar{r}}\,^*\right)^2\,,
\ee
and the bare $2k$ replica Hamiltonian
\be 
\quad\quad H_{0,k} \equiv \sum_{r=1}^k \left(H_0^{r} + H_0^{\bar{r}}\right)\,.
\ee 
For large $t,t'$ the dominant contribution to this partition function comes from the ground space and first excited states of $H_{\eff,k}$. The assumption, consistent with semiclassical and infinite-temperature (or large enough coupling $J$) considerations, is that the ground states break replica symmetry of $H_{\eff,k}$ and factorize into $k$ copies of the ground state of $H_{\eff,1}$ connecting two replicas. The two-boundary Euclidean wormhole which reproduces the ground state contribution is shown in Fig. \ref{fig:cosmo}. The first excited states correspond to single-particle excitations on each replica of this wormhole, with energy given by the two-replica gap $E_{\text{gap}}$. For the caterpillars of the SYK model this wormhole is a stabilized double-trumpet \cite{Magan:2024aet}.

\vspace{-.5cm}

This produces a late-time behavior
\be 
G_k(t,t') \approx k! F_1(\infty)^k \left(1 + k N_* e^{-\Egap(t+t')}\right)\,,
\ee 
where $k!$ comes from the number of ground states,  and $kN_*$ is the number of first excited states, corresponding to the possible single-particle excitations on each of the $k$ two-boundary wormholes. The late-time value is expressed in terms of the $k=1$ frame potential $F_1(\infty) = e^{-S_2(\rho)}$ for $\rho$ defined in \eqref{eq:equilibriumstate}. Using \eqref{eq:2norm}, this leads to a late-time distance to $k$-design
\be 
\Delta^\rho_{k}(t)^2 \approx N_*\,k!\,k\, e^{-2t \Egap}\,,
\ee 
which for large $k$ produces a time to $k$-design
\be\label{eq:lineargrowthsemi} 
t_{k} \approx \dfrac{1}{2} \Egap^{-1} \left(k\log k - k + \log k + \log N_{*} + 2\log \varepsilon^{-1}\right)\,.
\ee
Up to subleading logarithmic factors the growth is linear in $k$, with the slope $(2E_{\text{gap}})^{-1}$. The semiclassical analysis requires $k \ll e^{S_2(\rho)}$ to be able to neglect other non-perturbative effects. The length-randomness relation \eqref{eq:length_randomness} follows directly from \eqref{eq:lineargrowthsemi} together with \eqref{eq:length}. 

\textbf{Discussion} --- The ensembles of caterpillars constructed in this letter provide a window into the generic structure of the black hole Hilbert space in any theory of gravity with low-energy matter. The construction and main result of this letter support a vastly more general form of ER = EPR and seem to be in some tension with arguments against semiclassicality of typical interiors \cite{Almheiri:2012rt,Almheiri:2013hfa,Bousso:2012as,VanRaamsdonk:2013sza,Marolf:2013dba}. We believe that resolving this tension could be relevant for the typical-state firewall paradox. We think caterpillars can be useful for this purpose because, from the randomness perspective, we see no reason why they should have their properties dramatically altered before exponential times, even though we cannot show this with semiclassical methods. In future work, this extrapolation of the semiclassical analysis could be addressed in explicit UV complete models at finite $N$ using numerical or even experimental methods. Another open question is whether caterpillars offer any insight into weaker notions of firewalls in typical states, such as highly boosted matter shocks near the horizon \cite{Shenker:2013yza,Susskind:2015toa,Stanford:2022fdt,Blommaert:2024ftn,Iliesiu:2024cnh,Bousso:2025udh}. Naively, since the matter in the caterpillar begins at rest, the states do not seem to exhibit this type of firewall. Past-evolved caterpillars, evolved over a few scrambling times, do, however, because the matter becomes highly boosted in the frame of the infaller. At the level of ensembles of microscopic quantum states, both the caterpillars and the past-evolved caterpillars form approximate quantum state $k$-designs under the $2$-norm definition used in this paper. While, at the technical level, this points to the need for a stricter $1$-norm definition of an approximate $k$-design, it conceptually suggests that distinguishing states with firewalls from those without is extremely difficult. In fact, if we could extrapolate our result to exponentially large values of $k$, the indistinguishability would be so strong that the very question of whether a state has a firewall might become meaningless. It would be interesting to know whether this limitation is what the overlap analysis of \cite{Stanford:2022fdt,Blommaert:2024ftn,Iliesiu:2024cnh} is indicating. It would also be worth exploring whether the state-dependent interior reconstruction of \cite{Papadodimas:2012aq} plays any role here, given that our conclusion is largely state-independent.

Finally, in the spirit of the ``quantum gravity in the lab'' program~\cite{Brown_2023}, it would be interesting to construct and study these states in the lab as a way of directly probing the black hole interior. Physically instantiating these states is not trivial since they involve non-unitary elements due to the gradual cooling, but such evolutions can be implemented inefficiently using post-selection or a variety of other methods, e.g. the recently discussed double-bracket flow approach~\cite{gluza2024doublebracketquantumalgorithmsquantum}. Alternatively, by constructing an isometry which embeds the $e^S$ low-energy states of the black hole into the microscopic Hilbert space (e.g. as was effectively done in \cite{sewell2022thermalmultiscaleentanglementrenormalization} for a non-interacting fermion system), one might be able to combine this isometry with a conventional random circuit in the low-energy Hilbert space to produce random low-energy states embedded in the microscopic Hilbert space. \\

\textbf{Acknowledgments} --- We thank Stefano Antonini, Jos\'{e} Barb\'{o}n, Adam Bouland, Horacio Casini, Matt Headrick, Don Marolf, Douglas Stanford, Lenny Susskind, and Mark Van Raamsdonk for discussions. MS and BS acknowledge support from the U.S. Department of Energy through DE-SC0009986, QuantISED DE-SC0020360, and GeoFlow DE-SC0019380. The work of JM is supported by CONICET, Argentina. This preprint is assigned the code BR-TH-6724. 

\appendix
\renewcommand{\thesection}{\Alph{section}}

\section{A. Properties of typical equilibrium EPRs}
\label{app:states}

In this appendix, we study the ensemble of random purifications $\ens^\rho_{\text{EPR}}$ of a generic equilibrium density matrix $\rho$. The states have the form
\be\label{eq:randqubit2} 
\ket{\Psi^\rho_{\text{EPR}}} =  \dfrac{1}{\sqrt{Z_X}}\sum_{i,j} \sqrt{\rho_i \rho_j} \,X_{ij} \ket{E_i^*}_L\ket{E_j}_R\,.
\ee 
where $\ket{E_i}$ is an eigenstate of the density matrix $\rho$ with eigenvalue $\rho_i$ and $X_{ij}$ are Gaussian random coefficients with zero mean and unit variance. The quantity $Z_X = \, \text{Tr}(\rho X \rho X^\dagger)$ is a normalization factor. On average over the ensemble
\be
\mathbb{E}\left[Z_X\right]  =  \int \text{d}X\, \text{Tr}(\rho X \rho X^\dagger) =1\,.
\ee 
The variance, on the other hand, is 
\begin{gather}
\text{Var}\left[Z_X\right]  = \int \text{d}X \,\text{Tr}(\rho X \rho X^\dagger)^2-1= e^{-2S_2(\rho)} \,,
\end{gather} 
where $S_2(\rho) = -\log \text{Tr}(\rho^2)$ is the second R\'{e}nyi entropy. Since in our cases of interest $S_2(\rho)=O(N^2)$ we can forget about the normalization and consider ``annealed averages'' for the states normalized only on average.

The average reduced density matrices are given by the equilibrium state
\be 
\mathbb{E}\left[\rho_{L,R}\right] = \rho\,.
\ee 

Moreover, for individual instances, the $2$-norm distance to $\rho$ is 
exponentially suppressed,
\begin{align}
\text{Tr}(\rho_{L,R}^2) \approx \int \text{d}X \,\text{Tr}(\rho X \rho X^\dagger\rho X \rho X^\dagger)^2=\nonumber   2e^{-2S_2(\rho)} \,,
\end{align} 
From here, it directly follows that the $2$-norm distance is exponentially small,
\be 
\mathbb{E}\left[\big\|\rho_{L,R}- \rho\big\|_{2} \right] \leq \mathbb{E}\left[\big\|\rho_{L,R}- \rho\big\|^2_{2} \right]^{1/2} \approx e^{-S_2(\rho)}\,.
\ee 
Note, however, that the average $1$-norm distance to $\rho$ can be $O(1)$ so individual instances are distinguishable from the equilibrium state $\rho$.

\section{B. Exploration via time-independent vs. time-dependent Hamiltonians}

Let us focus on the $N$ qubit system. Consider time-evolving the TFD $\ket{I_t} = e^{-\iw t H_R}\ket{I}$, or explicitly,
\be\label{eq:inftempTFDtime} 
\ket{I_t} =    \dfrac{1}{\sqrt{D}}\sum_{i=1}^D e^{-\iw t E_i} \ket{E_i^*}_L\ket{E_i}_R\,.
\ee 

In order to compare this to the exploration via the random quantum circuit, we will diagonalize $U(t)$ and write the state \eqref{eq:inftempRC} as
\be\label{eq:randcqubitdiag} 
\ket{\Psi_t} =    \dfrac{1}{\sqrt{D}}\sum_{i=1}^D e^{-\iw \varphi_i(t)} \ket{\varphi_i(t)^*}_L\ket{\varphi_i(t)}_R\,,
\ee 
where $\ket{\varphi(t)_i}$ is an eigenvector of $U(t)$ with eigenvalue $\exp(-\iw \varphi_i(t))$ and $-\pi \leq \varphi_i(t)<\pi$. Finally, we will do the same thing for the typical EPR \eqref{eq:randqubit} and express it as
\be\label{eq:randqubitdiag} 
\ket{\Psi_{\text{EPR}}} =    \dfrac{1}{\sqrt{D}}\sum_{i=1}^D e^{-\iw \theta_i} \ket{\psi_i^*}_L\ket{\psi_i}_R\,,
\ee 
where $\ket{\psi_i}$ is an eigenvector of $U_{\text{Haar}}$ with eigenvalue $\exp(-\iw \theta_i)$ and $-\pi \leq \theta_i<\pi$. In this form, the structural similarity between \eqref{eq:inftempTFDtime}, \eqref{eq:randcqubitdiag} and \eqref{eq:randqubitdiag} is clear.

The eigenvectors $\ket{\psi_i}$ in \eqref{eq:randqubitdiag} are i.i.d. random vectors. The eigenphases $\theta_i$ are distributed according to the circular unitary ensemble (CUE) \cite{mehta2004random}
\be\label{eq:CUE} 
p_{\text{CUE}}(\theta_1,...,\theta_D) = \dfrac{1}{D!(2\pi)^D} \prod_{i<j} \left|e^{-\iw\theta_i}-e^{-\iw\theta_j}\right|^2\,.
\ee 
In the CUE, the eigenphases are uniformly distributed on average with a strong repulsion given by
\be\label{eq:2ptrepulsion}
\mathbb{E}_{\text{CUE}}[\delta \rho(\theta)\delta \rho(\theta')] \supset -\dfrac{\sin\left(D\,\frac{\theta - \theta'}{2}\right)^2}{\left(2\pi\,\sin\left(\frac{\theta - \theta'}{2}\right)\right)^2}\,.
\ee 
where $\rho(\theta) = \sum_{i=1}^D \delta(\theta-\theta_i)$ is the eigenphase density, and $\delta \rho(\theta) = \rho(\theta) - (2\pi)^{-1}$ is the fluctuation from the mean.

The question is whether the exploration via time-independent and time-dependent Hamiltonians is able to reproduce these statistical distributions of eigenphases/eigenvectors at sufficiently late times.

For time-independent Hamiltonians, there are two natural ensembles of states to be considered. One is to fix the Hamiltonian and consider the time evolution trajectory $\ens_{\text{erg}}= \lbrace \ket{I_t}: t \in \mathbf{R}\rbrace $ with uniform long-time measure $\lim_{T\rightarrow \infty}\frac{1}{T}\int_{-T/2}^{T/2}\text{dt}$. Any draw of $\ens_{\text{erg}}$ lives on a $D$ dimensional submanifold of $\mathscr{H}_{\text{EPR}}$, of torus topology, characterized by the phases in the energy eigenbasis \eqref{eq:inftempTFDtime} (we are neglecting that the global phase is unphysical). Assuming that the spectrum of the Hamiltonian lacks of rational relations, the ensemble $\ens_{\text{erg}}$ covers this set ergodically, and the long-time distribution of phases is the uniform measure on this torus
\be\label{eq:Poisson} 
p_{\text{erg}}(\theta_1,...,\theta_D) = \dfrac{1}{(2\pi)^D}\,.
\ee 
The eigenphases are independent and thus lack of repulsion -- the phase differences are Poisson distributed.

The second possibility is to define an ensemble at fixed time, $\ens^{H}_t$, by introducing disorder in the time-independent Hamiltonian $H$. Generally, the draws of the ensemble $\ens^{H}_t$ will explore the full $\mathscr{H}_{\text{EPR}}$ but will never become typical EPRs in $\ens_{\text{EPR}}$. In order to see this, consider a general spectral distribution for the eigenvalues of the Hamiltonian $E_i$, which may universally include eigenvalue repulsion. Essentially, without an extreme fine-tuning, at late enough times, the distribution of phases $\theta_i = t E_i\;(\text{mod }2\pi)$ of the entangled state \eqref{eq:inftempTFDtime} follows \eqref{eq:Poisson}. The reason is that, at timescales comparable with the mean level spacing of the Hamiltonian, the eigenphases $\exp(-\iw t E_i)$ will have wound around the unit circle so many times that all correlations between the $E_i$'s will be erased. Thus, no matter which of the two ensembles we consider, the exploration via time-independent Hamiltonians does not generate typical EPRs.

On the other hand, the ensemble $\ens_t$ generated with the random quantum circuit will have a statistical distribution of phases $\theta_i = \varphi_i(t)$, distributed according to $p_{t}(\theta_1,...,\theta_D)$ which, at infinite times, converges weakly to the CUE \cite{Guo:2024zmr}
\be 
p_{t}(\theta_1,...,\theta_D) \;\underset{t\rightarrow\infty}{\longrightarrow }\;p_{\text{CUE}}(\theta_1,...,\theta_D)\,.
\ee 
Likewise, $\ens_t$ will define a distribution of eigenvectors which converges weakly to the random state distribution.

All of these differences are quantified by the frame potential $F_k(\ens_\Psi) = \text{Tr}(\rho_k(\ens_\Psi)^2)$. For the ensemble of typical EPRs \cite{Jian:2022pvj,Guo:2024zmr,Magan:2024aet}
\be\label{eq:randFP} 
F_k(\ens_{\text{EPR}}) = \dfrac{k!}{D^{2k}}\quad\quad \text{for }k \leq D\,.
\ee 
The $k$-th frame potential is controlled by the $2k$-th moment of the CUE $\mathbb{E}_{\text{CUE}}[\rho(\theta_1)... \rho(\theta_{2k})]$. 

For the ensemble $\ens_{\text{erg}}$ the frame potential is a purely spectral quantity. It is controlled by the long time average of the $k$-th spectral form factor \cite{Cotler:2016fpe}, i.e., its ``plateau'' value, which for $k\leq D$ is
\be 
F_k(\ens_{\text{erg}}) =  \dfrac{1}{D^{2k}}\lim\limits_{T\rightarrow\infty}  \dfrac{1}{T}\int_{-T/2}^{T/2} \text{d}t \,\left|\text{Tr}(e^{-\iw t H}) \right|^{2k} = \dfrac{k!}{D^{k}}\,,
\ee
where in the last equality we have used the spectral ergodicity of $H$ to evaluate the long-time average. Already for the second moment ($k=1$) the frame potential differs from \eqref{eq:randFP} by factors of the dimension, which signals that the distribution of phases in $\ens_{\text{erg}}$ fails to incorporate the eigenphase repulsion \eqref{eq:2ptrepulsion}. A similar conclusion follows for $\ens_t^H$ at late enough times \cite{Cotler:2017jue}.

For $\ens_t$, the frame potential $F_k(t) \equiv F_k(\ens_t)$ is also a purely spectral quantity. In this case, it corresponds to the $2k$-th spectral form factor of the random quantum circuit, which at late times converges to the CUE value \cite{Jian:2022pvj,Guo:2024zmr,Magan:2024aet}
\be 
F_k(t) = \dfrac{1}{D^{2k}}\mathbb{E}| \text{Tr} U(t)|^{2k}\;\underset{t\rightarrow\infty}{\longrightarrow }\;F_k(\ens_{\text{EPR}})
\ee
The quantity $F_1(t)$ was in fact analyzed in \cite{Saad:2018bqo} as a generalization of the spectral form factor to a periodically driven Brownian SYK model.

\bibliographystyle{ourbst}
\bibliography{bibliography}

\end{document}